\documentclass[twocolumn]{el-author}

\usepackage{bbding}

\usepackage[misc]{ifsym}

\newcommand{\ua}{\uparrow}
\newcommand{\nc}{\newcommand}
\nc{\da}{\downarrow} \nc{\hc}{\hat{c}} \nc{\hS}{\hat{S}}
\nc{\bra}{\langle} \nc{\ket}{\rangle} \nc{\eq}{equation (\ref}
\nc{\h}{\hat} \nc{\hT}{\h{T}}\nc{\be}{\begin{eqnarray}}
\nc{\ee}{\end{eqnarray}}\nc{\rd}{\textrm{d}}\nc{\e}{eqnarray}\nc{\hR}{\hat{R}}\nc{\Tr}{\mathrm{Tr}}
\nc{\tS}{\tilde{S}}\nc{\tr}{\mathrm{tr}}\nc{\8}{\infty}\nc{\lgs}{\bra\ua,\phi|}\nc{\rgs}{|\ua,\phi\ket}
\nc{\hU}{\hat{U}}\nc{\lfs}{\bra\phi|}\nc{\rfs}{|\phi\ket}\nc{\hZ}{\hat{Z}}\nc{\hd}{\hat{d}}\nc{\mD}{\mathcal{D}}
\nc{\bd}{\bar{d}}\nc{\bc}{\bar{c}}\nc{\mc}{\mathcal}\nc{\ea}{eqnarray}\nc{\mG}{\mathcal{G}}\nc{\bce}{\begin{center}}
\nc{\ece}{\end{center}}
\date{XXth MONTH YEAR}

\begin{document}

\title{PUF-AES-PUF: a novel PUF architecture against non-invasive attacks}

\author{Weize Yu\small{\Envelope} and Jia Chen}

\abstract{

In this letter, a physical unclonable function (PUF)-advanced encryption standard (AES)-PUF is proposed as a new PUF architecture by embedding an AES cryptographic circuit between two conventional PUF circuits to conceal their challenge-to-response pairs (CRPs) against machine learning attacks.
Moreover, an internal confidential data is added to the secret key of the AES cryptographic circuit in the new PUF architecture to update the secret key in real-time against  side-channel attacks.
As shown in the results, even if 1 million number of  data are enabled by the adversary to implement machine learning or side-channel attacks, the proposed PUF can not be cracked.
By contrast, only 5,000 (1,000) number of data are sufficient to leak the confidential information of a conventional PUF via machine learning (side-channel) attacks.

}

\maketitle

\vspace{-0.1in}
\section{Introduction}
In the field of cybersecurity, silicon physical unclonable functions (PUFs) are well known hardware
security primitives which are applied in, for example, generating secret keys, executing authentications,
and devising wireless sensors [1].
The basic
mechanism of a silicon PUF is to exploit the intrinsic physical randomness of identically designed
integrated circuit (IC) components induced by the fabrication process to establish a unique mapping between the input challenges and the output responses (\textit{i.e.}, the challenge-to-response pairs
(CRPs)).

Although the present silicon PUFs may be effective for protecting the IoT against some specific
malicious attacks, like hardware reverse engineering attacks, one of the most significant
security concerns is their robustness against machine learning attacks [2].
High linear relationships exist between the CRPs for most conventional silicon PUFs, such as ring-oscillator
(RO) PUFs [3] and arbiter PUFs [3], which cause them to be extremely
vulnerable to machine learning.
All the improving techniques [2, 3, 4] to combat machine learning attacks are
focused on increasing the degree of non-linearity between the CRPs. Unfortunately, the high degree
of non-linearity may undermine the performance of silicon PUFs [2, 4].
Furthermore, improving the degree of non-linearity of silicon PUFs only hinders the execution of
machine learning attacks; it is incapable of thoroughly eliminating the threat of such attacks.

Side-channel attacks are a type of powerful non-invasive attacks that can be utilized by the
adversary to leak the confidential information of modern ICs through exploring the correlation between the processed data and the physical leakages (\textit{i.e.}, power dissipation, electromagnetic (EM)
emission, temperature, and timing) of the modern IC [5].
Advanced encryption standard
(AES) is a modern cryptographic algorithm that contains a stored secret key and is widely used
to encrypt confidential data [5]. It is well-known that the secret key of an unprotected AES
cryptographic circuit can be disclosed without much effort if side-channel attacks are implemented.
In order to secure an AES cryptographic circuit against side-channel attacks, all the existing countermeasures such as masking [5] and hiding [6, 7] are trying to break the aforementioned
correlation between the processed data and the physical leakages.
 Unfortunately, these existing
countermeasures [5, 6, 7] are not adequately secure and cause significant power/area/performance
overhead to the AES circuit.

In this letter, a novel and innovative PUF: key-updating (KU) AES-embedded PUF is devised against both machine learning and side-channel attacks, without increasing the degree of non-linearity between the CRPs and without using the existing countermeasures.

\begin{figure}[!t]
\centering
\vspace{-0.05in}
\includegraphics[width=3.6in]{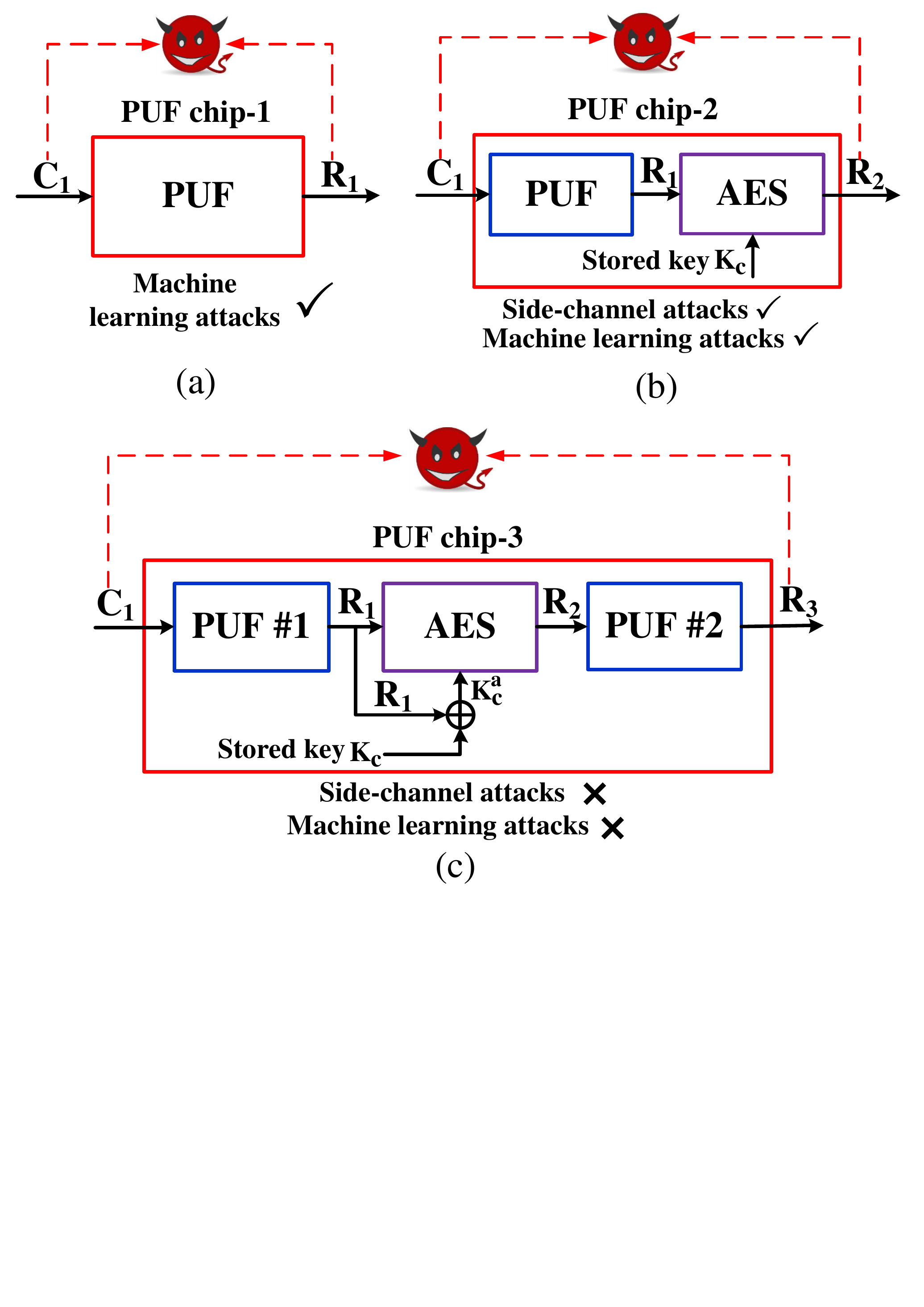}
\vspace{-1.9in}
\caption{Three different PUF chips under machine learning or side-channel attacks. (a) Conventional PUF (PUF chip-1). (b) Hybrid PUF (PUF chip-2). (c) Key-updating (KU) AES-embedded PUF (PUF chip-3).}
\vspace{-0.2in}
\label{1}
\end{figure}

\vspace{-0.1in}
\section{Working principle of the proposed PUF}
Fig. \ref{1}(a) shows a diagram of a conventional PUF chip under machine learning attacks. Since
both the input challenge $C_1$ and the output response $R_1$ are exposed to the adversary directly, 
the PUF chip-1 can be cracked by machine learning attacks through training a reasonable number of CRPs.
If we explore the design by connecting a conventional PUF circuit in series with an AES circuit to build a new PUF: hybrid PUF, as shown in Fig. \ref{1}(b),
the adversary may not be able to unriddle the secret information of the conventional PUF via machine learning attacks directly.
However, the hybrid PUF in Fig. \ref{1}(b) is not sufficiently secure.
The primary reason is that the output data $R_2$ of the AES is exposed to the adversary, therefore, the secret key $K_c$ of the AES
may be leaked to the adversary by analyzing the correlation between the output data $R_2$ and a
certain physical leakage of the PUF chip-2 if a side-channel attack is executed.
Once the secret key $K_c$ of the AES is leaked, the output response $R_1$ of the conventional PUF will also be disclosed.
As a result, the conventional PUF in Fig. \ref{1}(b) can be uncovered by training the $(C_1, R_1)$ pairs with
machine learning attacks ultimately.

So as to eliminate the threats from both side-channel and machine learning attacks, a KU AES-embedded PUF is proposed as shown in~Fig. \ref{1}(c).
The novel and innovative PUF architecture is secure against machine learning attacks, without increasing the degree of nonlinearity between the CRPs.
As indicated in~Fig. \ref{1}(c), an AES circuit is embedded between two conventional PUF circuits and the output response $R_1$ of the PUF \#1 circuit is encrypted by the
AES circuit to provide the input challenge $R_2$ to the PUF \#2 circuit.
Since the output response $R_1$ of the PUF \#1 and the input challenge $R_2$ of the PUF \#2 are concealed, the adversary is incapable
of performing machine learning attacks on either of the two PUF circuits.

Another novel innovation is proposed to eliminate the threat of side-channel attacks, which does
not rely on any of the existing countermeasures.
It is proposed to add a real-time key-updating
function to the architecture that combines the output response $R_1$ of the PUF \#1 with the stored
secret key $K_c$ of the AES to create the actual key $K_c^{a}$ used by the AES circuit ($K_c^{a} = R_1\oplus K_c$).
This is illustrated in~Fig. \ref{1}(c).
Since the input data $R_1$ and the output data $R_2$ of the AES are
unknown to the adversary and the actual secret key $K_c^{a}$ of the AES is updating in real-time, the
adversary is unable to execute side-channel attacks to discover the stored secret key $K_c$.

\begin{figure}[!t]
\centering
\vspace{0.1in}
\includegraphics[width=3.5in]{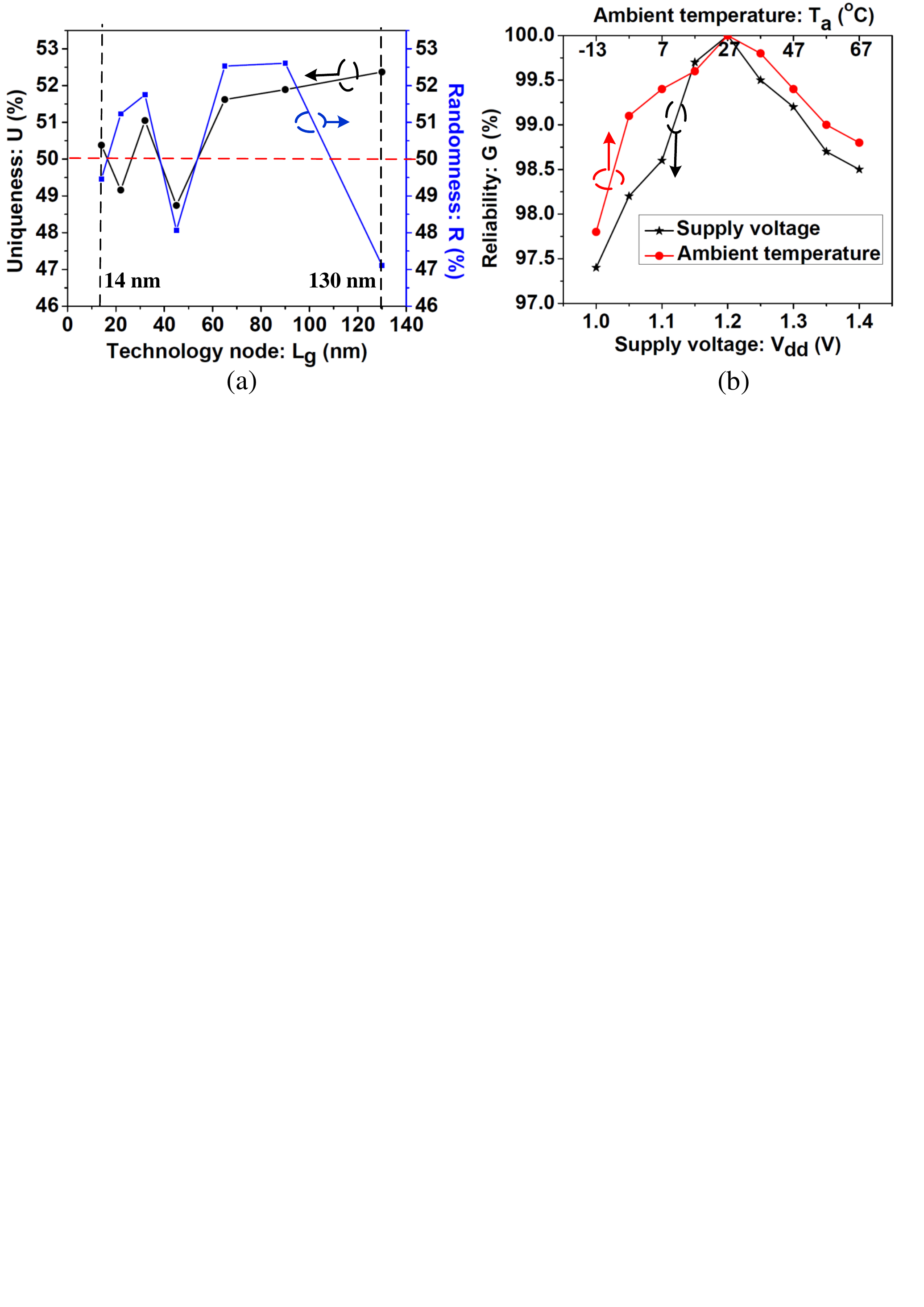}
\vspace{-3.72in}
\caption{Performance evaluation for the 128-bit KU AES-embedded PUF.
(a) Uniqueness $U$ and randomness $R$ versus technology node $L_g$. (b) Reliability $G$ versus supply
voltage $V_{dd}$ and ambient temperature $T_a$ ($L_g = 130$ nm).}
\vspace{-0.2in}
\label{2}
\end{figure}

\vspace{-0.1in}
\section{Performance evaluation}
Commonly, uniqueness, randomness, and reliability are the three most significant parameters for
assessing the performance of a designed PUF [1].
To evaluate the performance of the proposed PUF,
A 128-bit KU AES-embedded PUF is designed and simulated in Cadence software. 
The designed PUF consists of two 128-bit arbiter PUF circuits  (PUF \#1 and PUF \#2) and a 128-bit AES cryptographic circuit.
Moreover, Monte Carlo simulations are executed on the designed PUF in Cadence to emulate the random
fabrication process.
As shown in Fig. \ref{2}(a), the uniqueness $U$ is improved from 52.4\% to
50.4\% if the CMOS technology node $L_g$ is scaled from 130 nm to 14 nm; while the randomness $R$
improves from 47.1\% to 49.5\%. 
In addition, Fig. \ref{2}(b) shows the worst reliability of the embedded
PUF is about 97.4\% when the supply voltage is 1.0 V.
The simulation results manifest the
proposed PUF has excellent uniqueness, randomness, and reliability.

\begin{figure}[!t]
\centering
%\vspace{0.1in}
\includegraphics[width=3.45in]{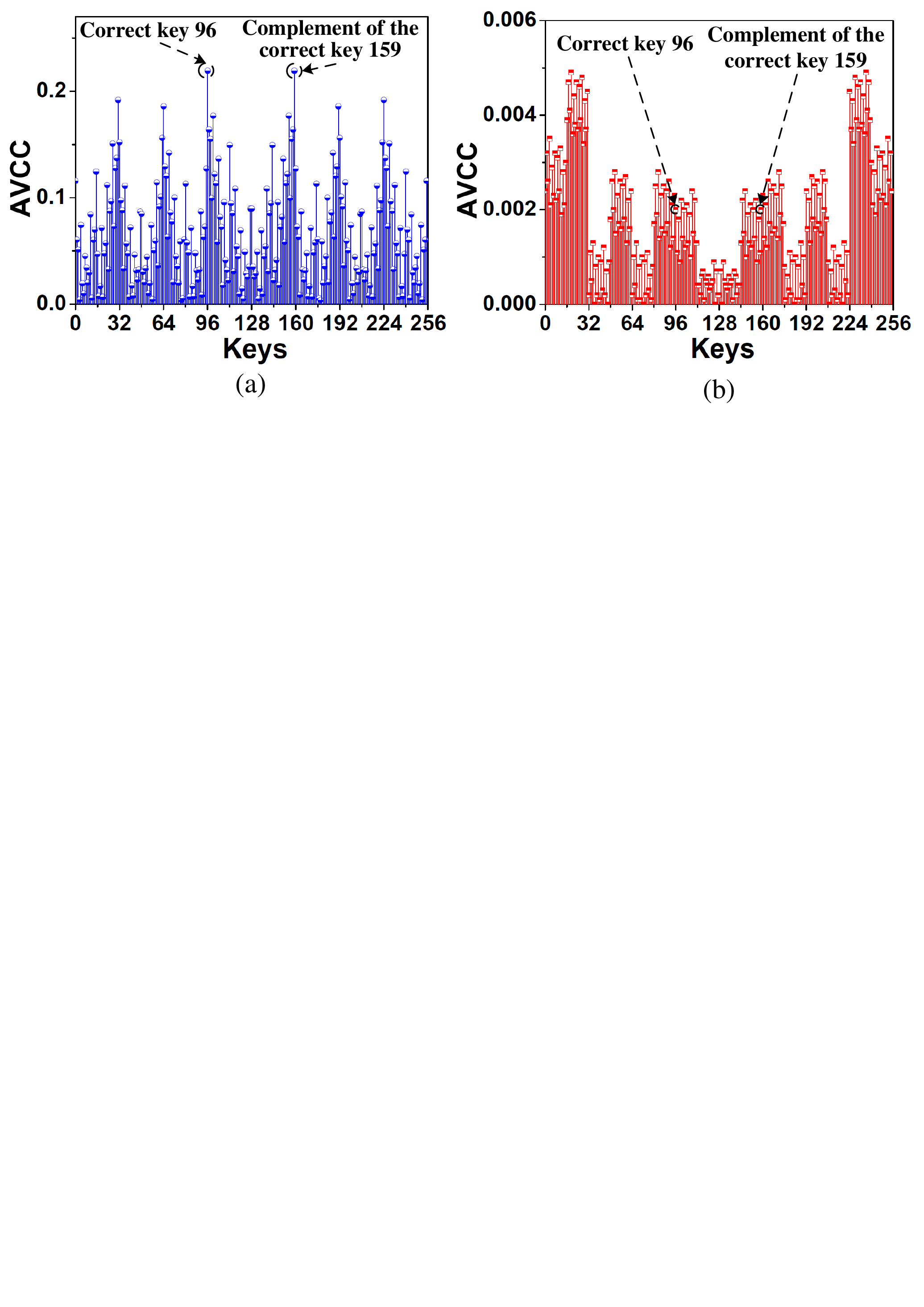}
\vspace{-3.64in}
\caption{Simulations of power attacks
(hamming-weight (HW) model is used).
(a) Absolute value of correlation coefficient (AVCC) versus possible keys for
leaking an 8-bit sub-key of the 128-bit unprotected AES cryptographic circuit after inputting 1,000 number of data. (b)
AVCC versus possible keys for leaking an
8-bit sub-key of the 128-bit KU AES-embedded PUF after inputting 1 million
number of data.}
\vspace{-0.2in}
\label{3}
\end{figure}

\vspace{-0.1in}
\section{Resilience against side-channel attacks} 
For the KU AES-embedded PUF we propose, the input data $R_1$ and
output data $R_2$ of the AES in Fig. \ref{1}(c) are unknown to
the adversary. As a result, to implement side-channel
attacks on the KU AES-embedded PUF, the adversary can only analyze the correlation between the input
challenge $C_1$/output response $R_3$ in Fig. \ref{1}(c) and the
physical leakages of the PUF chip.

Power attacks [6, 7] are a kind of side-channel attacks that
are widely used by the adversary to disclose the secret
key of a cryptographic circuit through monitoring the
correlation between the processed data and the power
dissipation of the cryptographic circuit. The
results of simulated power attacks are shown in Fig. \ref{3}
for (a) a 128-bit unprotected AES cryptographic circuit and (b) the 128-bit KU AES-embedded
PUF. As shown in Fig. \ref{3}(a), the 8-bit secret sub-key 96 of the unprotected AES circuit is
disclosed after inputting 1,000 plaintexts of data. However, for the AES-embedded PUF, the secret sub-key
96 is masked from being leaked to the adversary even if
1 million plaintexts are enabled, as shown in Fig. \ref{3}(b).
In addition, the absolute value of correlation coefficient
(AVCC) of the correct key 96 in Fig. \ref{3}(b) is two orders
of magnitude lower than the AVCC of the correct key
96 in Fig. \ref{3}(a). The primary reason is that the actual
secret key in the embedded PUF is updating in real-time which greatly weakens the correlation between the
processed data and the power dissipation against power
attacks.

\vspace{-0.1in}
\section{Robustness against machine learning attacks} 

Linear regression (LR) [3] and kernel support vector machine (SVM) [2] are two commonly used machine learning algorithms to estimate the 
relationship between the CRPs of a PUF.

Assume the input data of an  IC is an $n$-bit binary data $X=(x_1, x_2, ..., x_n)_2$ and the corresponding output data of the IC is an $m$-bit  
 binary data $Y=(y_1, y_2, ..., y_m)_2$.
If an LR algorithm is selected for executing a machine learning attack on the IC to model the \textit{1st} bit $y_1$ of  the output data $Y$, the relationship between the input data $X$ and the predicted output bit $y_1^{*}$ can be denoted as 
\begin{align}\label{eq:CCA}
&y_1^{*}=\sum_{i=1}^{n}a_ix_i+a_0
\end{align}
where $a_0$, $a_1$, ..., $a_n$ are the linear coefficients of the LR algorithm.
If $k$ number of input and output data pairs: ($X_1, Y_1$), ($X_2, Y_2$), ...,  ($X_k, Y_k$) are used as the training data sets, the cost function $F(a)$ of the LR algorithm and the coefficient $a_i, (i=1, 2, ..., n)$ under the gradient descend training can, respectively, be written as 
\begin{align}\label{eq:CCA}
&F(a)=\frac{1}{2k}\sum_{j=1}^{k}(\sum_{i=1}^{n}a_ix_{i,j}+a_0-y_{1,j})^2, \\
&a_i :=a_i-\beta\frac{\partial F(a)}{\partial a_i}=a_i-2\beta(\sum_{i=1}^{n}x_{i,j})F(a)
\end{align}
where $\beta$ is the learning rate of the LR algorithm
and
$x_{i,j}$ ($y_{1,j}$)  is the \textit{ith} (\textit{1st}) bit of the  \textit{jth}, ($j=1, 2, ..., k$) input (output) data $X_{j}$ ($Y_{j}$).

By contrast, if a kernel SVM algorithm is applied to study a machine learning attack on the IC, the classification function $G_{SVM}(X')$ is
\begin{align}\label{eq:CCA}
&G_{SVM}(X')=\mathrm{sign}(\sum_{j=1}^{k}\alpha_jY_{j}K(X_j,X')+b)
\end{align}
where $\alpha_j$ is the \textit{jth} support variable and $b$ is the optimal bias .
$K(X_j,X')$ denotes a Gauss kernel function whose bandwidth parameters are set as the median of the corresponding Euclidean distances.

\begin{figure}[!t]
\centering
\vspace{-0.18in}
\includegraphics[width=2.6in]{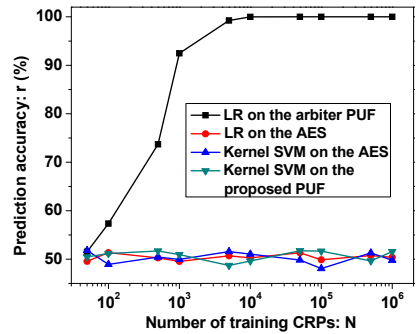}
\vspace{-0.2in}
\caption{Prediction accuracy $r$ versus number of training CRPs $N$ for the \textit{1st} output bit of
the three simulated circuits (designed with the 130 nm CMOS technology) : the 128-bit arbiter
PUF, the 128-bit AES, and the 128-bit proposed PUF under machine
learning attacks.}
\vspace{-0.2in}
\label{4}
\end{figure}

When the LR algorithm is used to train the CRPs of a 128-bit  simulated arbiter
PUF circuit, as shown in Fig. \ref{4}, the \textit{1st} output bit
of the 128-bit arbiter PUF circuit is cracked by the
LR algorithm after training about $5\times10^3$ number
of CRPs. By contrast, when the LR algorithm is
implemented on the 128-bit simulated AES cryptographic circuit, it fails to predict the output bit even if $1\times10^6$
number of CRPs are enabled for training. Moreover, as shown in Fig. \ref{4}, if a more advanced algorithm: kernel SVM is
selected for studying a machine learning attack, both the AES cryptographic circuit and the proposed PUF circuit are able to resist the kernel SVM. That indicates AES and AES-embedded
circuits are robust against the common machine learning attacks.

\vspace{-0.08in}
\section{Conclusion}
A novel  PUF-AES-PUF architecture in conjunction with a key-updating technique is utilized to design a state-of-the-art PUF primitive that is able to resist non-invasive attacks.
The proposed PUF not only has excellent uniqueness (52.4\%), randomness (47.1\%), and reliability (97.4\%) but also maintains a high security level (>1 million data) against both side-channel and 
machine learning attacks.

\vskip5pt

\noindent Weize Yu (\textit{Department of Electrical and Computer Engineering, Old Dominion University, Norfolk, VA 23529, USA})
\vskip3pt

\noindent  \small{\Envelope}  E-mail: wyu@odu.edu

\vskip5pt
\noindent Jia Chen (\textit{Department of Electrical and Computer Engineering, University of Minnesota Twin Cities, Minneapolis, MN 55455, USA})

\end{document}